\documentclass[12pt]{article}
\usepackage{times}
\usepackage{graphicx}
\usepackage{ulem}
\usepackage[usenames,dvipsnames]{color}
\usepackage[labelfont=bf,labelsep=period]{caption}

\newenvironment{sciabstract}{%
\begin{quote} \bf}
{\end{quote}}

\newcounter{lastnote}

\title{Direct observation of how the heavy fermion state develops in  \boldmath{CeCoIn$_{5}$} }

\author{Q.~Y.~Chen$^{1,2}$,  D.~F.~Xu$^{1}$, X.~H.~Niu$^{1}$,  J.~Jiang$^{1}$,  R.~Peng$^{1}$,  H.~C.~Xu$^{1}$,   \\
C.~H.~P.~Wen$^{1}$, Z.~F.~Ding$^{1}$,  K.~Huang$^{1}$,  L. Shu$^{1,7}$,  Y.~J.~Zhang$^{3,4}$,  H.~Lee$^{3}$,  \\
 V.~N.~Strocov$^{5}$,  M.~Shi$^{5}$,  F. Bisti$^{5}$, T.~Schmitt$^{5}$, Y. B. Huang$^{6}$ , P.~Dudin$^{7}$ \\
 X.~C.~Lai$^{2}$, S. Kirchner$^{3,9 \dag}$,   H.~Q.~Yuan$^{3,4,8}$,  and D.~L.~Feng$^{1,8 \ast}$
\\
\normalsize{$^1$State Key Laboratory of Surface Physics and Department of Physics},\\ \normalsize{Fudan University, Shanghai 200433, China}\\
\normalsize{$^2$Science and Technology on Surface Physics and Chemistry Laboratory},\\ \normalsize{Mianyang, 621908, China}\\
\normalsize{$^3$ Center for Correlated Matter, Zhejiang University, Hangzhou, 310058, China}\\
\normalsize{$^4$Department of Physics, Zhejiang University, Hangzhou, 310027, China}\\
\normalsize{$^5$Swiss Light Source, Paul Scherrer Institute, CH-5232 Villigen PSI, Switzerland}\\
\normalsize{$^6$Shanghai Institute of Applied Physics, CAS, Shanghai, 201204, China}\\
\normalsize{$^7$Diamond Light Source, Harwell Science and Innovation Campus, Didcot OX11 0DE, United Kingdom}\\
\normalsize{$^8$Collaborative Innovation Center of Advanced Microstructures, Nanjing 210093, China}\\
\normalsize{$^9$Department of Physics and Astronomy, Rice University, Houston, Texas, 77005, USA}\\
\\
\normalsize{E-mail: $^\ast$dlfeng@fudan.edu.cn, $^\dag$stefan.kirchner@correlated-matter.com }
}

\begin{document}
%\linenumbers

\baselineskip24pt


\maketitle
% * <dlfeng@fudan.edu.cn> 2016-08-11T06:26:25.215Z:
%
% ^.
% * <dlfeng@fudan.edu.cn> 2016-08-11T06:26:25.038Z:
%
% ^.


\begin{sciabstract}
Heavy fermion materials gain high electronic masses and expand  Fermi surfaces when the high-temperature localized $f$ electrons become itinerant and hybridize with the conduction band at low temperatures. However, despite the common application of this model, direct microscopic verification remains lacking.
 Here we report high-resolution angle-resolved photoemission spectroscopy measurements on CeCoIn$_5$, a prototypical heavy fermion compound, and reveal the long-sought band hybridization and Fermi surface expansion. Unexpectedly, the localized-to-itinerant transition occurs at surprisingly high temperatures, yet $f$ electrons are still largely localized at the lowest temperature.  Moreover,
crystal field excitations likely play an important role in the anomalous temperature dependence.
Our results paint an  comprehensive unanticipated experimental picture of the heavy fermion formation in a periodic multi-level Anderson/Kondo lattice, and set the stage for understanding the emergent properties in related materials.

%\sout{We report high resolution angle-resolved photoemission measurements on CeCoIn$_5$, a prototypical Heavy Fermion system. Upon cooling into the `coherent'  state, the $f$ electrons hybridize with the conduction electrons, develop heavy mass, and enlarge the Fermi surface. Our direct observation of this electronic weight gain reveals several surprises. In particular, the paradigmatic localized-to-itinerant $f$-electron transition develops at remarkably high temperatures, and exhibits an anomalous energy/temperature scaling behavior, in which crystal field excitations may play a substantial role.  The observed electronic structure evolution provides a comprehensive characterization of the heavy fermion state and its formation in CeCoIn$_5$ in unprecedented detail.   Our results paint a comprehensive experimental picture of a periodic multi-level Anderson/Kondo lattice system, and set the stage for understanding its unconventional superconductivity, magnetism and quantum phase transitions.}
\end{sciabstract}
\newpage
%
%{\bf One Sentence Summary:}\\
%\bl{We paint a comprehensive experimental picture of the development of heavy electrons in a periodic multi-level Kondo lattice.}
%Photoemission spectroscopy unfolds an unanticipated experimental picture of the development of heavy electrons in a periodic multi-level Kondo lattice.
%\bl{We paint a comprehensive experimental picture of the development of electron obesity in a periodic multi-level Kondo lattice.}
% Wasn't quite brave enough to do that unilaterally on someone else's paper.
% --Darren

%\section{\label{sec:level1}General Introduction}

%While the initial interest was centered around the Fermi liquid properties and the occurrence of unconventional superconductivity, recent research has been focused on the general phase diagram of this material class and possible underlying quantum critical points.
The comparatively high tunability of the heavy fermions in Cerium- and Ytterbium-based rare earth intermetallics makes them attractive systems in the search for emergent phases of matter. This tunability is partially due to the dynamic generation of low energy scales and the concomitant enhancement of the density of states at the Fermi energy ($E_F$).
%The emergence of  low energy scale in a heavy fermion system is  manifested in its  behaviors as  a function of temperature.
According to the standard model of heavy fermion behavior, epitomized by the periodic Anderson model in its local moment regime \cite{Varma.76}, the $f$ electrons are localized at high temperatures, while their hybridization with conduction electrons leads to the formation of bands with heavy masses as temperature is lowered and the $f$-electrons become  itinerant\cite{Stewart.84}. Exotic superconductivity, quantum criticality, and magnetic order subsequently emerge from these heavy electrons.

Such a generic picture of the localized-to-itinerant transition has been demonstrated by dynamical mean field calculations\cite{Shim.07,Choi.12}, which predict a dramatic change of the Fermi surface due to the inclusion of fully itinerant $f$ electrons. However, the experimental picture has been puzzling, if not unclear, and direct validation is still lacking \cite{Mo.12,Kummer.15,Fujimori.07}.
Moreover, previous experimental studies mainly focused on the low-temperature properties, and could not trace how the heavy-electron state is formed through the localized-to-itinerant transition.
We take CeCoIn$_5$ (superconducting transition temperature $\sim$~2.3~K) as an example, since much of our current knowledge on heavy fermion systems comes from the studies of this compound.
De~Haas-van~Alphen (dHvA)~\cite{Settai.01,Shishido.02}, optical conductivity\cite{Singley.02,Burch.07}, and scanning tunneling microscopy (STM) \cite{Aynajian.12,Allan.13,Zhou.13} measurements compare well with band calculations that assume fully itinerant $f$ electrons \cite{Shishido.02,Haule.10}, but they cannot provide the detailed electronic structure evolution with a sufficient temperature range and momentum resolution.
Curious angle-resolved photoemission spectroscopy (ARPES) results  \cite{Koitzsch.13,Koitzsch.08,Koitzsch.09} have been obtained for CeCoIn$_5$: some suggest the Ce 4$f$ electrons are itinerant up to 105 K \cite{Koitzsch.08}, while later results suggest they are predominantly localized at 25 K \cite{Koitzsch.09}.
Moreover, a recent ARPES study on YbRh$_2$Si$_2$ between 90\,K and 1\,K even challenged this picture --- the anticipated Fermi surface change is not found in this temperature range\cite{Kummer.15}. (see, however, Ref. \cite{Paschen.16}).

The appearance of various temperature scales in different measurements further complicates the problem. For example, with decreasing temperature, the resistivity of CeCoIn$_5$ first increases and then starts decreasing just below 50 K\cite{Petrovic.01}. This characteristic temperature for the resistivity maximum, $T_{coh}$, has been associated with the onset of a coherent heavy electron band, and assumed to be close to the Kondo temperature, $T_K$. Meanwhile, STM found a scaling of the local conductance in CeCoIn$_5$ between 20 and 60~K~\cite{Aynajian.12}. Intriguingly, both the Seebeck and the Nernst coefficients exhibit anomalies at 20 K \cite{Bel.04}.
How the  heavy fermion state develops with such perplexing multiple characteristic temperatures, and the role of any other similar energy scale besides $T_K$, {\it e.g.}\ crystal field excitations, remain matters of debate.

Here, we address these perplexing findings in the prototypical heavy fermion system CeCoIn$_5$. We combine bulk-sensitive soft x-ray ARPES to unravel its three-dimensional (3D) electronic structure, and resonant ARPES to expose the $f$-electron behavior in an extended temperature range with much improved energy resolution.
As a result, a comprehensive experimental picture of heavy fermion formation unfolds with unprecedented precision and details. Our data  give a microscopic verification of the current model, and help to understand various  experimental results. More importantly, we  discovered  unexpected important behaviors beyond the current model, which further deepens our  understanding of heavy fermion systems.

%These findings may help build a comprehensive experimental picture of heavy fermion systems in general.

\section*{3D Fermi surface and band structure}

%%%%%%%%%%%%%%%%%%%%%%%%%%%%%%%%%%%%%%%%%%%%%%%%%%%%%%%%%%%%%%%%Fig1

\begin{figure}
\centering
\includegraphics[width=16cm]{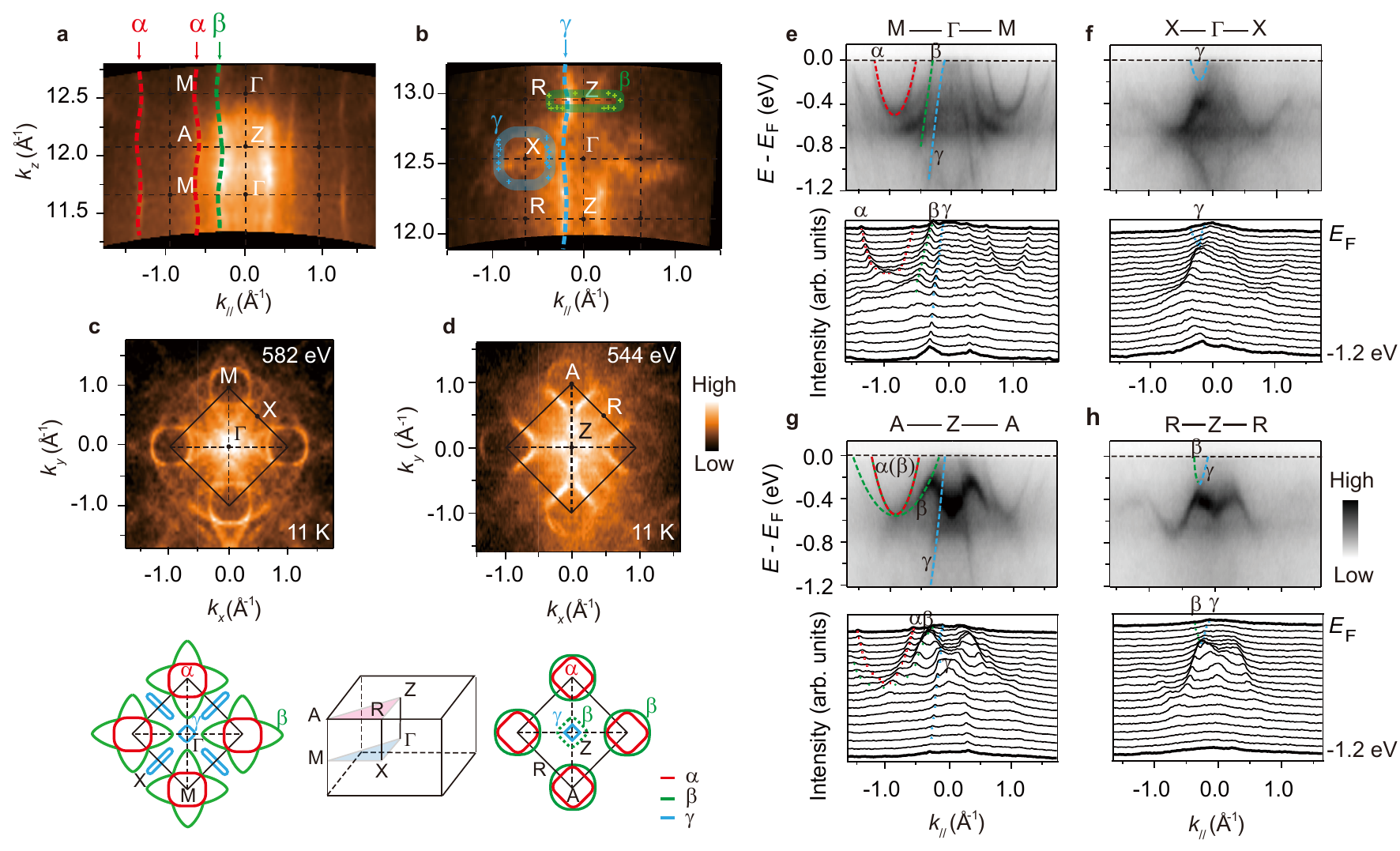}
\caption{{\bf Three dimensional electronic structure of CeCoIn$_5$ at 11 K.}
\textbf{a},\textbf{b}, Photoemission intensity maps in the $\Gamma$$ZAM$ (\textbf{a}) and $\Gamma$$ZRX$ plane (\textbf{b}), respectively. Different $k_z$s were accessed by varying the photon energy between 480 and 648~eV. \textbf{c},\textbf{d}, Photoemission intensity maps in the $\Gamma$$XM$ (\textbf{c}) and $ZAR$ (\textbf{d}) plane taken with 582 and 544 eV photons respectively, which were determined based on the periodicity of the high symmetry planes.
The inner potential was estimated as 16~eV.
 The lower panels plot the Fermi surface sheets in the upper panels by tracking Fermi crossings.  Part of the $\beta$ Fermi surface is hard to trace (dashed lines). The  Brillouin zone of CeCoIn$_5$ is depicted in the middle.   \textbf{e-h}, Photoemission intensity distributions along $\Gamma$-$M$ (\textbf{e}) and $\Gamma$-$X$ (\textbf{f}), $Z$-$A$ (\textbf{g}) and $Z$-$R$ (\textbf{h}), respectively. Lower panels are their corresponding momentum distribution curves (MDCs).
All photoemission intensity data here were integrated over a window of ($E_F$-20 meV, $E_F$+20 meV).
 }
\end{figure}
%%%%%%%%%%%%%%%%%%%%%%%%%%%%%%%%%%%%%%%%%%%%%%%%%%%%%%%%%%%%%%%%%%%

We start by measuring the 3D electronic structure of CeCoIn$_5$ at 11~K with soft x-ray ARPES, whose high out-of-plane momentum ($k_z$) resolution makes it an effective tool for studying more-3D rare-earth compounds \cite{Strocov.12}.  Photoemission intensity maps in the $\Gamma ZAM$ plane and $\Gamma ZRX$ plane of the CeCoIn$_5$ Brillouin zone are shown in Figs.~1a and 1b, respectively.

The Fermi surface in the $\Gamma$$XM$ plane, shown in Fig.~1c, is composed of two electron pockets --- a flower-shaped $\beta$ and a rounded $\alpha$ pocket --- and two hole pockets --- a square-like pocket around $\Gamma$ and a narrow racetrack-shaped pocket around $X$, both attributed to the $\gamma$ band (Fig.~1f). As shown by the photoemission intensity plot along the $\Gamma$-$M$ direction in Fig.~1e, the $\alpha$ band is parabolic-like with its bottom 0.45 eV below $E_F$. In the $ZAR$ plane (Fig.~1d), the $\beta$ pocket becomes rounded and the $\alpha$ pocket square-like. As shown in Fig.~1g, the $\beta$ band contributes two Fermi crossings along $Z$-$A$, one degenerates with the $\alpha$ band while the other contributes part of the square-like Fermi surface around $Z$.

Our data show that the $\alpha$ band is the most two-dimensional (2D). Although the cross-sections of the $\alpha$ and $\beta$ Fermi surfaces show weak variation in the $\Gamma$$ZAM$ plane (Fig.~1a), obvious differences can be observed between the Fermi surface topologies in the $\Gamma$$XM$ and $ZAR$ planes. In the $\Gamma$$ZRX$ cross-section (Fig. 1b), there is a small $\beta$-derived pocket around $Z$ and another pocket around $X$ contributed by the $\gamma$ band, indicating rather 3D characters of these two bands. The shapes of the $\alpha$ and $\beta$ pockets qualitatively agree with previous dHvA measurements and  calculations \cite{Settai.01,Shishido.02}. While dHvA provided little information on the $\gamma$ band, we find that it contributes a square-like pocket around $Z$ in the $ZAR$ plane, contradicting the calculation which predicts that the $\gamma$ Fermi surface does not cross the $ZAR$ plane \cite{Shishido.02}. The difference may arise from the only partially-itinerant $f$ electrons, as will be shown later in this paper. The observed quasi-2D electronic structure is consistent with the moderate in-plane vs.\ out-of-plane anisotropies in the superconducting critical field and magnetic susceptibility \cite{Petrovic.01,Shishido.02,Pfleiderer.09}.

\section*{On- and off- resonance ARPES data}

%%%%%%%%%%%%%%%%%%%%%%%%%%%%% Figure. 2 %%%%%%%%%%%%%%%%%%%%%%%%%%%%
\begin{figure}
\includegraphics[width=12cm]{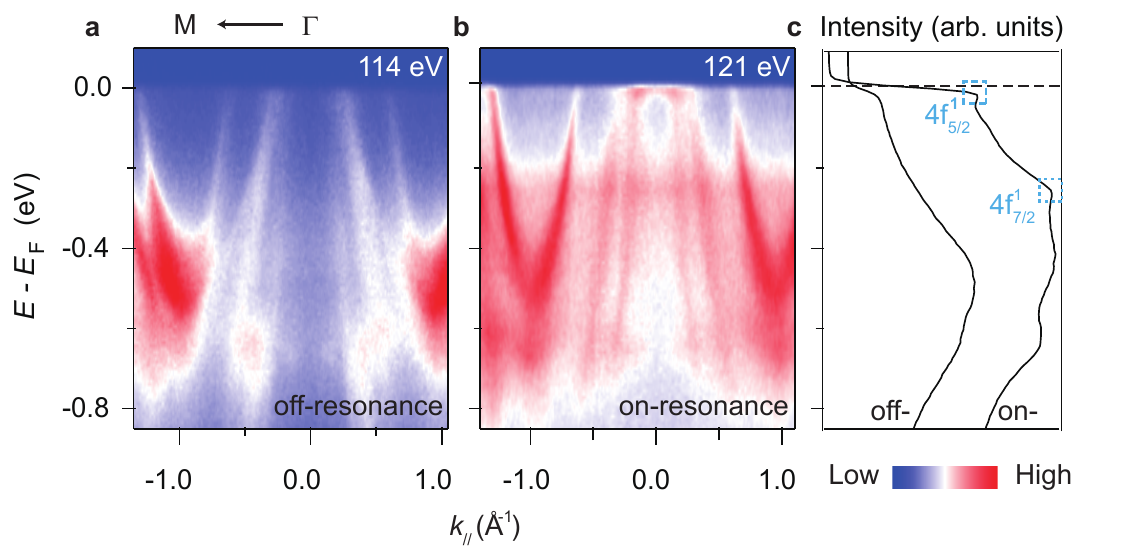}
\centering
\caption{
\textbf{On- and off- ARPES data of CeCoIn$_5$ at 11 K.}
\textbf{a},\textbf{b}, Photoemission intensity distributions of CeCoIn$_5$ taken along  $\Gamma$-M with off- (114 eV) and on-resonance (121~eV) photons, respectively, at 17 K.  \textbf{c}, The angle-integrated EDCs taken with off- and on-resonant energy; $f$ band peak positions are highlighted. The labels follow the common notation in Refs.~\cite{Im.08,Fujimori.03,Fujimori.06}. The momentum cut taken with 121~eV photons crosses (0,0,7.08~$2\pi/c$), close to $\Gamma$, and is thus labeled as  $\Gamma$-M for simplicity.
}
\end{figure}
%%%%%%%%%%%%%%%%%%%%%%%%%%%%%%%%%%%%%%%%%%%%%%%%%%%
To enhance the $f$ electron photoemission matrix element, resonant ARPES measurements were conducted at the Ce 4$d$-4$f$ transition. Figs.~2a and 2b show the photoemission intensities taken off-resonance and on-resonance at 17 K, respectively. Emission from Co 3$d$ and Ce 5$d$ states dominates the off-resonance spectra, while Ce 4$f$ emission is enhanced in the on-resonance data, as also shown by the integrated spectra in Fig. 2c. Two nearly flat features can be observed in the on-resonance data, corresponding to the $4f_{5/2}^{1}$ and $4f_{7/2}^{1}$ states\cite{Fujimori.03,Fujimori.06}. The feature near $E_F$ is actually the tail of the Kondo resonance peak above $E_F$.

\section*{Temperature dependence of the \boldmath{$f$}  electrons}

Figure.~3a shows the temperature-evolution of the resonant ARPES data along $\Gamma$-$M$. At high temperatures, the photoemission intensity is dominated by the strongly dispersive $d$ bands. Upon decreasing temperature, two weakly dispersive $f$-electron features near $E_F$ and -280~meV gradually emerge.
The temperature dependence can be further demonstrated by the energy distribution curves (EDCs) at $\Gamma$ in Fig.~3b. To access features above $E_F$, the spectra are divided by the resolution-convoluted Fermi-Dirac distribution (RC-FDD) at corresponding temperatures and shown in Fig.~3c.
One can identify three features located at 2, 9, and 30~meV above $E_F$, labeled as $E_1$, $E_2$ and $E_3$, respectively. The $E_1$ peak spectral weight gradually increases with decreasing temperature from 190~K to 17~K (Fig.~3d, and caption).
The temperature dependencies of the peaks at $E_2$ and $E_3$ appear non-monotonic --- they are most prominent at 60 and 120~K, respectively.

%%%%%%%%%%%%%%%%%%%%%%% Figure 2 %%%%%%%%%%%%%%%%%%%%%%%%%%%%%%%%
\begin{figure}
\includegraphics[width=\linewidth]{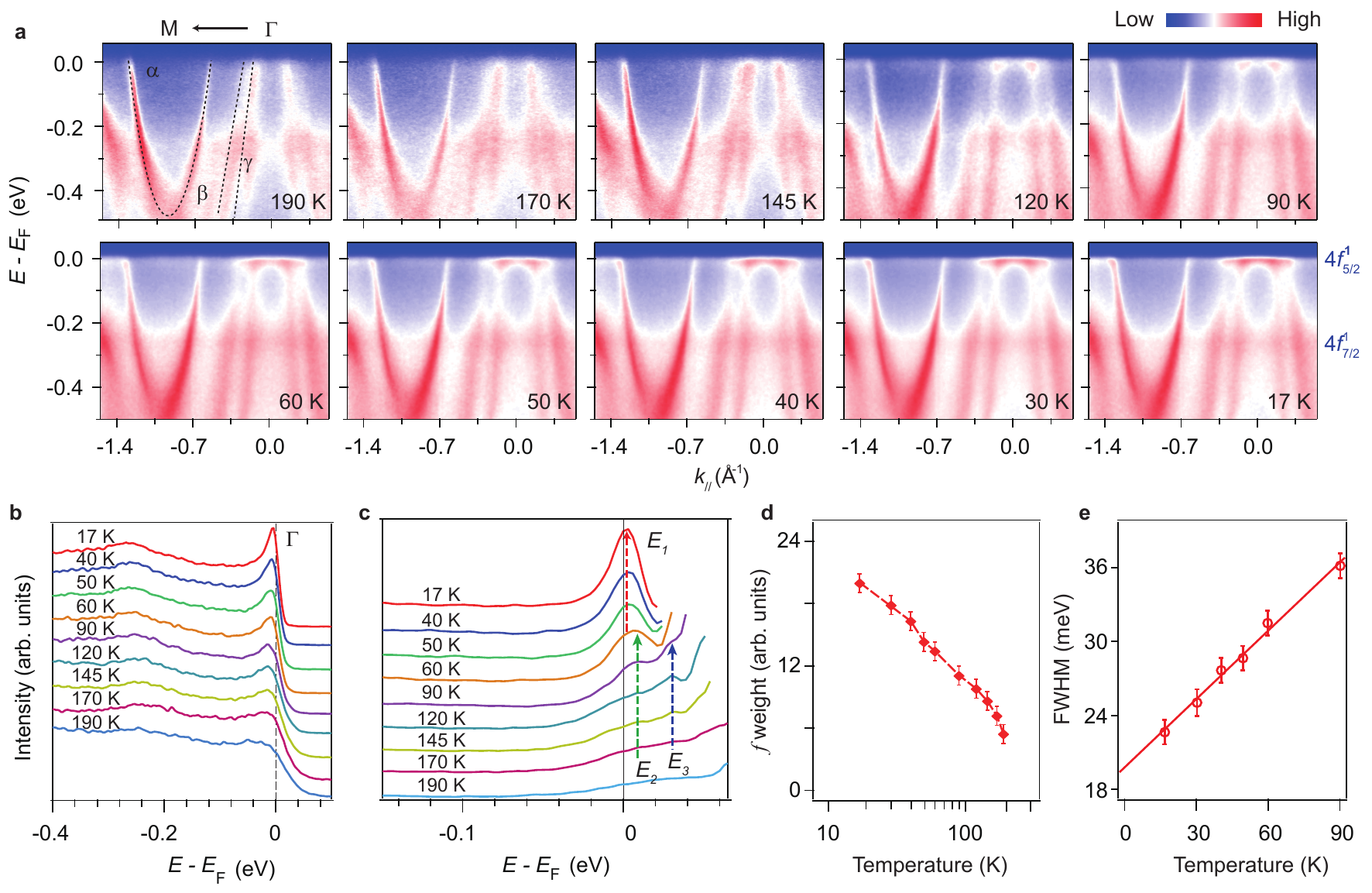}
\caption{
\textbf{Temperature evolution of the electronic structure of CeCoIn$_5$.}
\textbf{a}, Resonant (121~eV) ARPES data of CeCoIn$_5$ along  $\Gamma$-$M$ at the labeled temperatures.
 The momentum cut taken with 121~eV photons crosses (0,0,7.08~$\frac{2\pi}{c}$), close to $\Gamma$, thus labeled hereafter as  $\Gamma$-M for simplicity.
\textbf{b}, Temperature dependence of the EDCs at $\Gamma$. \textbf{c}, To reveal states above $E_F$, the spectra in (\textbf{b}) are divided by the RC-FDD at their respective temperatures. Three features at $E_1$=2~meV, $E_2$=9~meV and $E_3$=30~meV can be observed.
\textbf{d}, The background-subtracted integrated quasiparticle spectral weight near $E_F$ in the vicinity of $\Gamma$ as a function of temperature on log-linear scale. This is calculated by integrating the left half of the main peak in (\textbf{c}) over [-40~meV, 2~meV] after subtracting the flat background over [-140~meV, -98~meV]. \textbf{e}, Temperature dependence of the FWHM of the  main Kondo resonance  at $E_1$ in (\textbf{c})  below 90~K (circles). The data are obtained after correcting for the influence from the first excited CEF state, following the analysis in  Fig. S1 of Supplementary. The line is the best fit,  giving  FWHM [meV]= 20 + 2.15$\times k_B T$~[K].
}
\label{fig:Tdep}
\end{figure}
%%%%%%%%%%%%%%%%%%%%%%%%%%%%%%%%%%%%%%%%%%%%%%%%%%%%%%%%%%%%%%%%%%

For CeCoIn$_5$, one expects the six-fold degenerate $4f_{5/2}$ state to be split into three Kramer's doublets by the tetragonal crystal electric field (CEF). Each excited CEF doublet can participate in Kondo scattering processes and give rise to Kondo resonance satellite peaks shifted from the main Kondo resonance \cite{Kroha.03}. The peak separations, $E_2-E_1\sim$ 7~meV, and $E_3-E_1\sim$ 28~meV, are in excellent agreement with the CEF splittings estimated by neutron scattering \cite{Bauer.04}. Therefore, we conclude that the features found here are crystal-field-split $4f_{5/2}$ states, with the main Kondo resonance peak at $E_1$ above $E_F$, as expected for a Ce-based heavy fermion system \cite{footnote1}.

It is commonly believed that the localized-to-itinerant transition sets in around $T_{coh}$, but our results demonstrate that the heavy band formation begins at much higher temperatures than previously conceived. For example, Fig.~3a shows that the feature near $E_F$ is already clearly discernible around 120~K, and Fig.~3d  shows that the $f$-electron weight increases upon cooling from the highest measured temperature  (190~K). In fact, the extrapolation conducted in Fig.~S2 of Supplementary suggests an onset around 270$\pm$30~K.
It is commonly accepted that  the CEF  splittings can enhance the Kondo temperature when the CEF level separations and Kondo energy scale are comparable~\cite{Cornut.72,Kroha.03}.
While it has been difficult to unambiguously identify CEF excitations in thermodynamic and transport measurements, particularly for CeCoIn$_5$,
our observation of Kondo satellite peaks (Fig. 3c) supports the analysis of the CEF excitations of Bauer \textit{et al.}~\cite{Bauer.04}, and  suggest that the CEF excitations are most likely responsible for the observed high onset temperature of the transition in CeCoIn$_5$.

Figure.~3e plots the single-particle scattering rate of the main Kondo resonance, represented by the full-width at half-maximum (FWHM) of the peak at $E_1$ in Fig.~3c. Intriguingly, it shows a linear temperature dependence for $T\leq90$\,K, and the zero-temperature width of 20~meV, obtained through extrapolating the fitting lines, is largely resolution-limited.
A $T$-linear scattering rate has been observed in many unconventional superconductors \cite{Daou.09,Dai.13,Bruin.13}, most notably the optimally-doped cuprates\cite{Valla.99}.

Figures 4a and 4b compare the photoemission intensity maps near $E_F$ taken at 170 K and 17 K, respectively. Clearly, the $f$ spectral weight is much enhanced around $\Gamma$ at low temperature, indicating that the $f$-electrons participate in the Fermi surface formation.
Figs.~4c and 4d give a detailed comparison. At 170~K, the bands exhibit almost linear dispersion with large slopes, while at 17~K, all three bands are bent by hybridization with the flat $f$ band that becomes coherent at low temperatures, and they become weakly dispersive near $E_F$. Figs.~4e and 4f zoom into the vicinity of the Fermi crossing of $\alpha$ at 60~K and 17~K respectively, where the  data have been divided by the corresponding RC-FDDs. Because of the influence of the first excited CEF state, the dispersion above $E_F$  cannot be precisely traced in the 60~K data, however, one can clearly observe the Fermi velocity of $\alpha$ decreasing when going from 60 to 17~K.  Compared to the high temperature dispersion in Fig. 4c (also see Supplementary Fig. S3), the Fermi velocity of $\alpha$ is reduced by a factor of $24 \pm 3$ at 17 K.  The $\gamma$ band  behaves similarly, as shown in Figs. 4g-h. We note that due to the overlap of several bands, the hybridization of $\beta$ cannot be resolved, but it is expected to behave similarly to $\alpha$.

%%%%%%%%%%%%%%%%%%%%%%%%%%%% Figure 3 %%%%%%%%%%%%%%%%%%%%%%%%%%%%
\begin{figure}
\includegraphics[width=\linewidth]{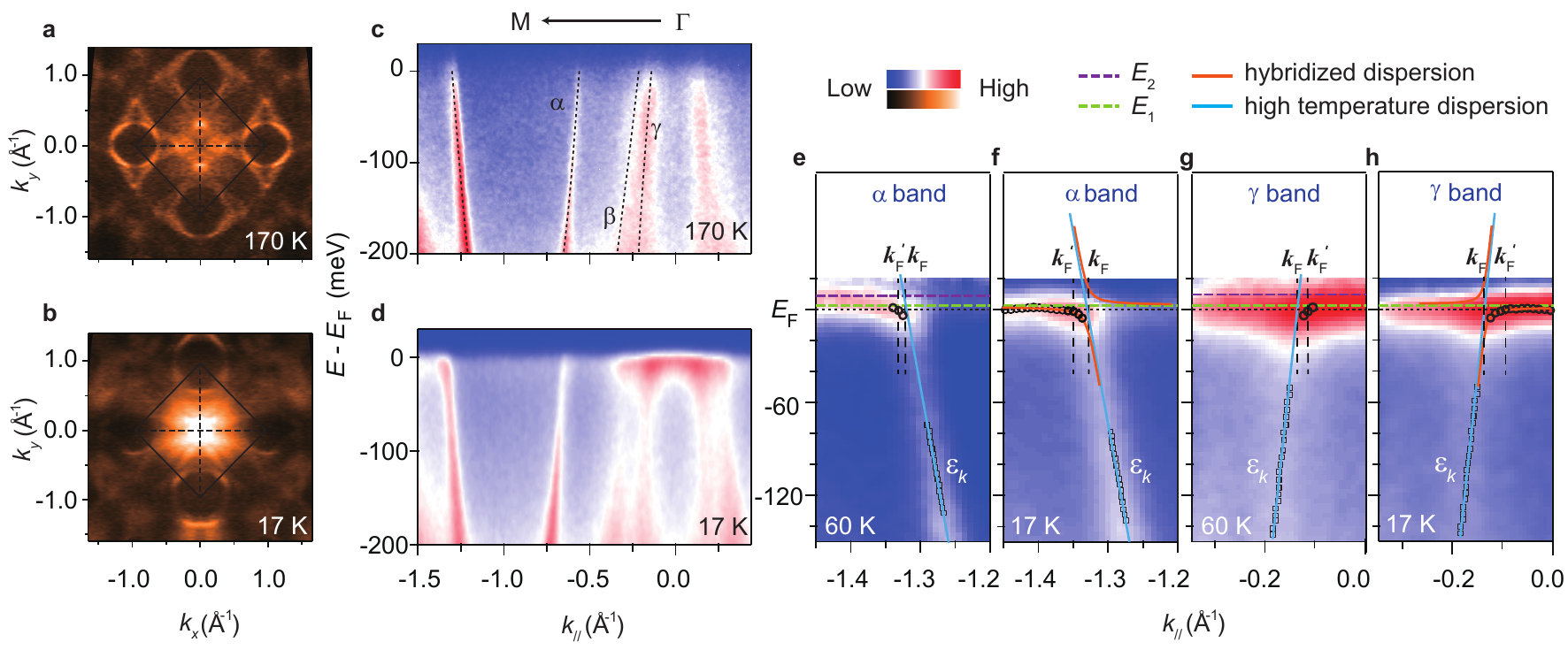}
\caption{
\textbf{Development of heavy electron states upon cooling.}
\textbf{a},\textbf{b}, Photoemission intensity maps with 121 eV photons at 170 K (\textbf{a}) and 17 K (\textbf{b}).  \textbf{c},\textbf{d}, Photoemission intensity distributions along $\Gamma$-$M$ taken at 170 K (\textbf{c}) and 17 K. \textbf{d},\textbf{f}, Photoemission intensity distribution  near the Fermi crossing of the $\alpha$ band taken at 60~K (\textbf{e}) and 17 K (\textbf{f}) after divided by the RC-FDD. \textbf{g},\textbf{h}, same as panels (\textbf{e}) and (\textbf{f}) except they are near the Fermi crossing of the  $\gamma$ band. Circles represent the position of the hybridized $f$ band obtained by tracking EDCs, squares represent the position of the conduction band at high temperature from fitting MDCs. The $f$-$d$ hybridizations in panels (\textbf{f}) and (\textbf{h}) are modeled by the periodic Anderson model (Eq. 1, orange curves), where the blue lines denote the high-temperature band dispersions. $E_1$ and $E_2$ are indicated by dashed lines.
}
\end{figure}
%%%%%%%%%%%%%%%%%%%%%%%%%%%%%%%%%%%%%%%%%%%%%%%%%%%%%%%%%%%%%%%%%%

% Following paragraph:
% I've removed ", where the effective masses range from 9 to 20 $m_e$ in the normal state" (dHvA)
% I hope this wasn't important
% --Darren

When the $f$ electrons are localized at high temperatures,  the effective masses of the $\alpha$ and $\gamma$ bands can be obtained with the parabolic fits shown in Supplementary Fig.~S4 of their dispersions over a several hundred meV window, which give
$m_{\alpha}^{*}$ = 0.88~$m_e$, and $m_{\gamma}^{*}$ = 0.24~$m_e$ ($m_e$ being the free electron mass), as labeled respectively.
Based on the Fermi velocity renormalization, $m_{\alpha}^{*}$ is thus estimated to be 21.1 $m_e$ at 17 K, consistent with dHvA \cite{Hall.01}. For the  $\gamma$ band, its Fermi velocity is renormalized by $102 \pm 10$ times at 17~K (Supplementary Fig. S3), which gives a effective mass $m_{\gamma}^{*}$ of  24.5 $m_e$ at 17 K.

%The formation of heavy-electron states at low temperature is illustrated in Fig.~3G, where the effective masses of both the $\alpha$ and $\gamma$ bands exhibit $\ln(T)$ dependence. This typical Kondo behavior is demonstrated directly by ARPES for the first time.

Phenomenologically, although the system is not a simple Fermi liquid, our 17~K data can be  well described by a simple mean-field hybridization band picture based on the periodic Anderson model (PAM) \cite{Hewson.93} as presented in Figs.~4f and 4h, in which the energy dispersion is given by
\begin{equation}
E^{\pm}=\frac{\varepsilon_0+\varepsilon(k)\pm\sqrt{(\varepsilon_0-\varepsilon(k))^2+4|V_k|^2}}{2},
\end{equation}
\noindent where $\varepsilon_0$ is the renormalized $f$-level energy (the CEF ground state here), $\varepsilon_k$ is the conduction-band dispersion at high temperatures (Fig.~S3), and  $V_k$ is the renormalized hybridization \cite{Im.08,Hewson.93}.
A fit to this model gives $\varepsilon_0$=2~meV and $V_k$=15~meV for both $\alpha$ and $\gamma$, respectively, corresponding to a direct gap (defined as the minimal separation of two bands at the same momentum) of 30\,meV. Therefore, the indirect gap is vanishingly small, if not zero (defined as the minimal separation of two bands in energy). The previous picture of a sizable indirect gap proposed by the STM study (and the hybridization model therein) is an overestimate. Because the tunneling matrix element for the conduction band is much larger than that of the $f$ electrons \cite{Aynajian.12}, STM mainly measures the direct gap, like optical conductivity \cite{Singley.02}.

In either Figs.~4e-f (for $\alpha$) or Figs.~4g-h (for $\gamma$), the Fermi crossing ($k_F'$) departs further from the high temperature position ($k_F$) when the temperature is lowered from  60~K to 17~K,  clearly demonstrating the gradual inclusion of additional $f$ electrons into the Fermi sea upon cooling into the heavy electron state.
Based on the quantitative analysis in Supplementary Fig. S5, we estimate that  the radius of the  $\alpha$ electron pocket \textit{expands} by $0.024\pm0.008$\AA$^{-1}$ between 145~K and 17~K,
while the radius of the  $\gamma$ hole pocket \textit{shrinks} by $0.042\pm0.008$\AA$^{-1}$  in this temperature range.
Considering the rather two-dimensional character of  $\alpha$ and assuming an isotropic expansion, we estimate that 0.058 electrons have been transferred from the local $f$ moment to the $\alpha$ pocket at low temperature. Similarly,  we estimate that 0.032 holes are removed from the central $\gamma$ pocket.
Since the Fermi surfaces in Figs.~4a and 4b do not show any other pronounced difference,
we loosely estimate the total Fermi surface volume expansion to be about $0.1\pm0.02$ electrons in the localized-to-itinerant transition,  consistent with the Ce valence measured by core-level x-ray absorption \cite{Booth.11} and photoemission \cite{Treske.14}.
Our data indicate that the $f$-electrons only become partially itinerant above 17~K, in contrast to various theories and calculations suggesting fully-itinerant $f$-electrons at such moderately low temperatures and the dramatic change of Fermi surface topology  \cite{Choi.12}.

\section*{Intermittent \boldmath{$E/T$}  scaling and CEF excitations}
%%%%%%%%%%%%%%%%%%%%%%%%%%%%% Figure 4 %%%%%%%%%%%%%%%%%%%%%%%%%%%%
\begin{figure}
\centering
\includegraphics[width=7.5cm]{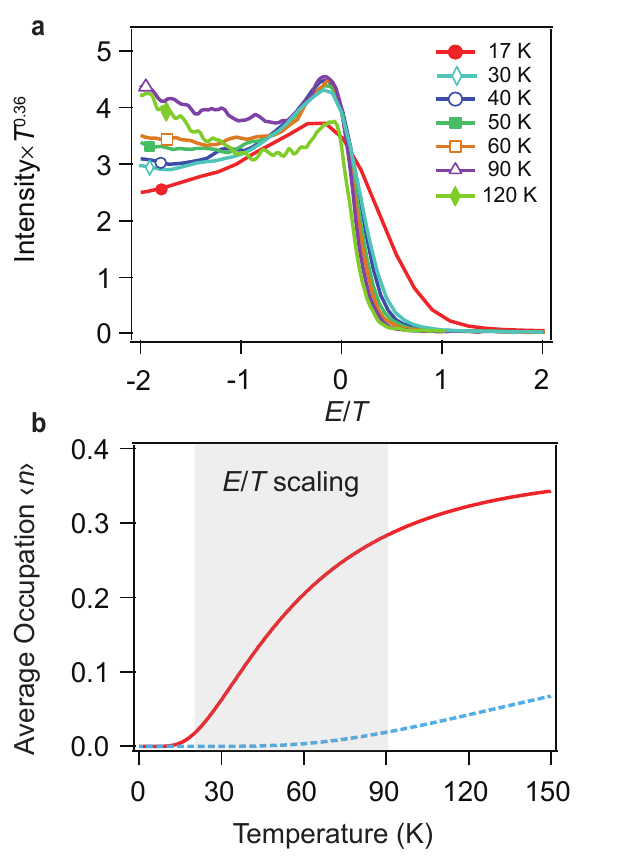}
\caption{
\textbf{Intermittent $E/T$ scaling and the CEF excitation occupations}.
\textbf{a},
The EDCs at $\Gamma$  are scaled by $T^{0.36}$ and plotted over $E/T$ for individual temperatures. In this way, the $E/T$ scaling can be fulfilled for data taken between 90 and 30 K.
\textbf{b},  Estimate of the average occupation of the first (red solid line) and second (blue dashed line) excited CEF doublets as a function of temperature assuming CEF splittings of 7 meV and 28 meV, see Supplementary for details. The gray region denotes the temperature range with $E/T$ scaling.
}
\label{fig:QC}
\end{figure}
%%%%%%%%%%%%%%%%%%%%%%%%%%%%%%%%%%%%%%%%%%%%%%%%%%%%%%%%%%%%%%%%%%
The resistivity of CeCoIn$_5$ shows a maximum at $T_{coh} \approx 50$~K and decreases almost linearly upon cooling down to the 2.3\,K onset of superconductivity, and the quasiparticle scattering rate also exhibits a linear dependence on temperature (Fig. 3e).
The $T$-linear scattering rate in the cuprates has been interpreted in terms of a phenomenological marginal Fermi liquid theory and an underlying energy-over-temperature ($E/T$) scaling of the single-particle excitations that is associated with quantum-critical fluctuations \cite{Varma.89,Varma.16}.
$E/T$ scaling and concomitant linear inverse lifetimes have also  been  established in a number of quantum critical heavy fermion compounds \cite{Aronson.95,Schroder1,Friedemann.10}. Such a `quantum critical' scaling of the local conductance with a concomitant linear-in-$T$ broadening has been reported in a recent STM experiment on CeCoIn$_5$ to set in below  60 K \cite{Aynajian.12}.

Our ARPES data obey $E/T$ scaling with a fractional exponent $\approx 0.36$ between 90 and 30 K as well (Fig.~5a) \cite{footnote2}. Notably, the 17 K and 120~K data fail to show scaling.
The absence of $E/T$ scaling below roughly 20 K indicates that it does not underlie the linear-in-temperature behavior of the resistivity at these temperatures \cite{footnote3}.
%\textbf{I suggest to delete these following. Stefan please revise, I think maybe we just leave them out, and focus on the observations and direct suggestions}
%In-line with Theoretical considerations. While CeCoIn$_5$ is located in proximity to an anti-ferromagnetic QCP, it is commonly believed that this QCP is of the spin-density wave or Hertz-Moriya-Millis type, where $E/T$ scaling with a fractional exponent is not expected.  Our results call for further theoretical investigations. For example, it is interesting to check if and how the $E/T$ scaling observed here relates to the quantum critical fluctuation spectrum of a Kondo-destroying QCP~\cite{Si2001}, which occurs in a sister compound, CeRhIn$_5$.
Instead, the anomalous  temperature-dependent scaling seems to be related to the depopulation of the excited CEF levels in CeCoIn$_5$.
%maybe understood after considering the CEF effects, since
%CEF splittings imply a $T$-dependence of occupation, which would introduce additional temperature scales observed during the Kondo lattice formation.
Fig.\ 5b shows the anticipated occupation of the CEF-split $4f_{5/2}^{1}$ state, based on the obtained CEF splittings of 7 meV and 28 meV (see also Supplementary):
below 20~K, essentially all $4f_{5/2}^{1}$ electrons occupy the lowest CEF doublet; above 20 K, the occupation of the first excited state increases with temperature, undergoing a smooth slope change around 90~K and leveling off slowly at higher temperature. This suggests that the various marked changes below 20~K could be linked to the depletion of the excited CEF levels, including the absence of  $E/T$ scaling of the spectral function, the slope change of the Seebeck coefficient at 20 K, which commonly signals lattice coherent Kondo scattering, and the strong temperature and field dependence of the Nernst coefficient below 20 K \cite{Bel.04}.
In addition, the onset of $E/T$ scaling as temperature is lowered coincides with
the increase in the depopulation rate of the excited levels as temperature drops below the lowest CEF gap around 90 K.

%The temperature dependence thus mimics that of Fig. 5B.
%The resistivity maximum occurs just below 50~K, {\it i.e.} in a temperature window where the spectral function near $E_F$ around the $\Gamma$ point displays $E/T$ scaling.

Understanding the emergence of unconventional superconductivity and quantum criticality in heavy fermions is linked to an understanding of the low-energy scales in these systems~\cite{Stockert.12}.
Our results strongly suggest that CEF effects play an unexpectedly significant role in the localized-to-itinerant transition of CeCoIn$_5$. The scope and significance of their involvement in the formation of the heavy fermion state have so far not been fully appreciated. They will not only enhance the Kondo temperature but also introduce additional characteristic temperatures, and may be ubiquitous to all the Ce-based rare earth intermetallics.

In conclusion, our results directly paint an experimental picture of the localized-to-itinerant transition in a prototypical periodic Kondo lattice system.
Besides providing the first 3D Fermi surface mapping and experimental band structure of the heavy fermion compound CeCoIn$_5$, we show how the localized $f$ electrons become partially itinerant and evolve into the heavy fermion state from much higher temperatures than $T_{coh}\approx$ 50~K, and how the Fermi volume increases along the way. The observed temperature dependence of the electronic structure unambiguously indicates the importance of CEF excitations. Thus, our results  provide
 a nearly complete microscopic picture of how the heavy fermions in CeCoIn$_5$ develop and evolve with temperature, which will be essential for building a microscopic theory of the  unconventional superconductivity and quantum phase transitions  in this compound.

 % The  critical facts and quantitative knowledge found here also  for  heavy fermion materials.

\subsection*{Methods}

Single crystals of CeCoIn$_5$ were grown by an In self-flux method. Room-temperature powder x-ray diffraction measurements revealed that all the crystals are single-phase and crystallize in the tetragonal HoCoGa$_5$ structure. The samples were then cleaved along the $c$-axis in ultra-high vacuum before performing ARPES measurements.
Soft X-ray ARPES data shown in Fig.~1 were taken at the Advanced Resonant Spectroscopies (ADRESS) beamline at the Swiss Light Source, with a variable photon energy and PHOIBOS-150 analyzer. The overall energy resolution was 70-80 meV, and the angular resolution was 0.07$^{\circ}$. The samples were cleaved and measured at 11 K under a vacuum better than $5\times 10^{-11}$~mbar.
All the  data taken with the on-resonance 121 eV photons  (except those in Figs. 3a-b) were obtained at the ``Dreamline" beamline of the Shanghai Synchrotron Radiation Facility (SSRF) with a Scienta D80 analyzer. The samples were cleaved $in$ $situ$ at 17 K. The vacuum was better than $5\times10^{-11}$~mbar at 17~K. The energy resolution was 17 meV, and the angular resolution was $0.2^{\circ}$.
The data in Figs. 3a-b were taken at Beamline I05-ARPES of the Diamond Light Source, equipped with a Scienta R4000 analyzer. The typical angular resolution was $0.2^{\circ}$ and the overall energy resolution was better than 16 meV. The vacuum was kept below $9\times10^{-11}$~mbar. The samples were cleaved at 170 K before performing ARPES experiments. All the data were taken with $p$-polarized light.

\section*{Acknowledgments}
We gratefully acknowledge enlightening discussions with Prof.\ Y.\ F.\ Yang, J.\ X.\ Zhu,
and the experimental support of Dr.\ M.\ Hoesch, Dr.\ T.\ Kim,  Ms.\ X.\ P.\ Shen,
and Dr.\ Y.\ J.\ Yan. This work is supported in part by the National Science Foundation of
China (Grant No. 11504342, U1630248), National Key R\&D Program of the MOST of China (Grant No. 2016YFA0300200),
Science Challenge Program of China, Science and Technology Commission of Shanghai Municipality (Grant No. 15ZR1402900), and Diamond Light Source for time on beamline I05
under Proposal No. SI11914.
The soft x-ray ARPES experiment has been performed at the SX-ARPES endstation of the ADRESS beamline at the Swiss Light Source, Paul Scherrer Institute, Switzerland. F.\ Bisti is supported by the Swiss National Science Foundation.

\section*{Author contributions}
Q.C., D.X., X.N., J.J., H.X. and C.W. performed ARPES measurements, Z.D., K.H., L.S., Y.Z., H.L. and H.Y. grew single crystals, V.S., M.S., F.B., T.S., Y.H. and P.D. supported the synchrotron experiments, Q.C., R.P., S.K. and D.F. analyzed the data. D.F. and S.K. wrote the paper, D.F. and X.L. are responsible for the infrastructure, project direction and planning.

\newpage
\clearpage
\setcounter{figure}{0}
\renewcommand{\thefigure}{S\arabic{figure}}
\setcounter{equation}{0}
\renewcommand{\theequation}{S.\arabic{equation}}
\setcounter{section}{0}
\renewcommand{\thesection}{S.\Roman{section}}
\renewcommand{\thesubsection}{S.\Roman{section}.\Alph{subsection}}
\makeatletter

\section*{Supplementary material for: Direct observation of how the heavy fermion state develops in  \boldmath{CeCoIn$_{5}$}  }

{\centering\subsection*{Abstract}}

In this supplementary material, we present simulation of the peak-width correction (Fig. S1), determination of the onset temperature of the localized-to-itinerant transition (Fig. S2), linear fits of the band dispersions near $E_F$ (Fig. S3), parabolic fit of the large scale dispersion at 170 K (Fig. S4), and quantitative analysis of the enlargement of the $\alpha$ and $\gamma$ bands (Fig. S5). We also present detailed statements of the ``CEF excitations and the temperature dependence of the CEF level occupation" and ``Implications for YbRh$_2$Si$_2$".

\subsection*{Simulation of the peak-width correction}

%%%%%%%%%%%%%%%%%%%%%%%%%%%%% Figure. S2 %%%%%%%%%%%%%%%%%%%%%%%%%%%%
\begin{figure}[ht]
\centering
\includegraphics[width=\textwidth]{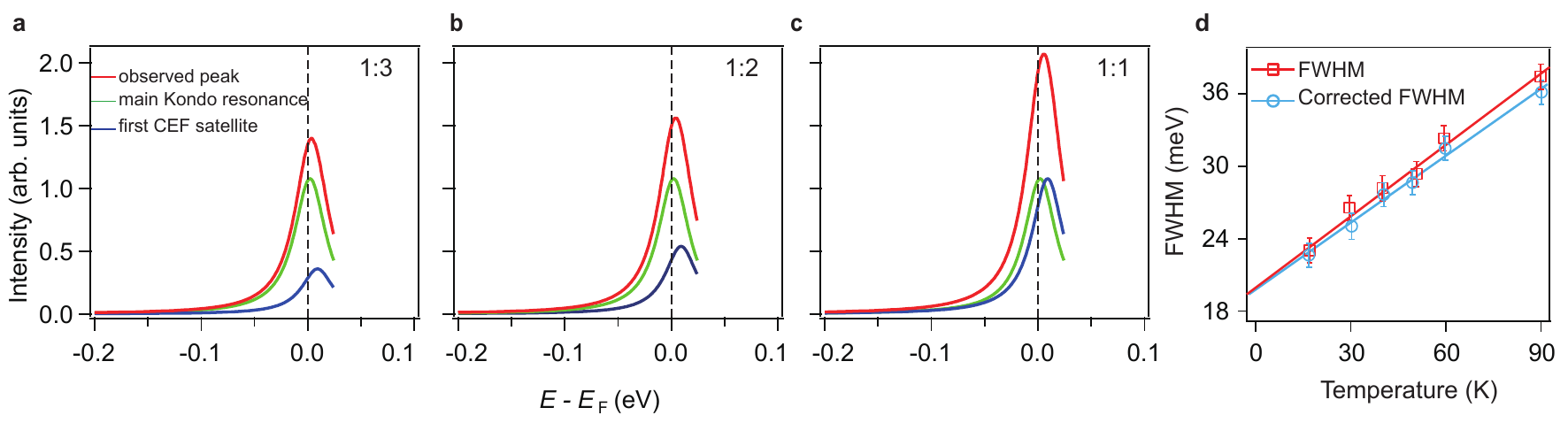}
\caption{
\textbf{a}-\textbf{c}, Simulation of the influence of the first CEF satellite to the FWHM of the main Kondo resonance peak. The green curve denotes the main Kondo resonance peak, which is located about 2 meV above $E_F$. The blue curve is the first CEF satellite (assuming the same width), which is 7 meV higher. The red curve is the sum of the two peaks. The ratio of the intensity of the first CEF excitation to the Kondo resonance peak is 1$:$3 in (\textbf{a}),  1$:$2 in (\textbf{b}) and 1$:$1 in (\textbf{c}), respectively. \textbf{d}, Temperature dependence of the FWHM of the main peak in Fig. 3c   below 90~K (squares). It is determined from twice the left half-width at half-maximum of this peak. Since this has contributions from both the ground and first excited CEF states, the FWHM of the CEF ground state (circles) is obtained after correcting for the minor influence from the first excited CEF state. The best fit lines are FWHM [meV]= 20 + 2.29$\times k_B T$~[K] (red), and FWHM [meV]= 20 + 2.15$\times k_B T$~[K] (blue), respectively.
}
\end{figure}
%%%%%%%%%%%%%%%%%%%%%%%%%%%%%%%%%%%%%%%%%%%%%%%%%%%%%%%%%%%%%%%%%%

The main peak observed in Fig.\ 3c has contributions from both the main Kondo resonance peak (the ground CEF state) and the first CEF excitation satellite. To estimate the influence of the first CEF excitation to the FWHM of the main Kondo resonance peak, we performed a simulation by considering the main Kondo resonance peak located at 2 meV above $E_F$ and the first CEF excitation satellite located 7 meV higher. We assume that they have the same width, which takes the experimental value.
When the intensity ratio between the first CEF excitation and the main Kondo resonance peak is 1:3, 1:2 and 1:1, we found that the FWHM measured from twice the left half-width at half-maximum of the combined peak is only 1.3, 1.7, and 2.1 meV larger than that of the main Kondo resonance peak, respectively (see Fig. S1).   This illustrates that the experimental FWHM estimated  in this way largely reflects the FWHM of the main Kondo resonance peak.

This also provides a way to obtain a more accurate FWHM of the main Kondo resonance.  Fig. S1d shows both  the temperature dependence of  the FWHM directly measured from the main peak in Fig.~3c, and that of the corrected FWHM for the main Kondo resonance. The latter is also presented in Fig.~3e.

\newpage
\subsection*{Determination of the onset temperature of the localized-to-itinerant transition}
%%%%%%%%%%%%%%%%%%%%%%%%%%%%% Figure. S3 %%%%%%%%%%%%%%%%%%%%%%%%%%%%
\begin{figure}[ht]
\centering
\includegraphics[width=7.5cm]{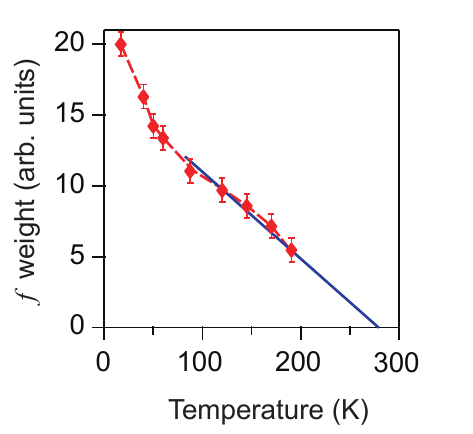}
\caption{
The temperature dependence of  the $f$ electron spectral weight (data shown in Fig.~3d) plotted in a linear-linear scale. The linear
extrapolation to high temperatures indicates that the $f$ electron spectral weight reaches zero around 270 K.
}
\end{figure}
%%%%%%%%%%%%%%%%%%%%%%%%%%%%%%%%%%%%%%%%%%%%%%%%%%%%%%%%%%%%%%%%%%

The  heavy-electron states develop with decreasing temperature, enhancing the $f$-electron spectral weight (Fig.~3d).
Such a process started from the highest measured temperature. To estimate the onset of the localized-to-itinerant transition, we re-plot the data in Fig.~3d on a linear-linear scale and  extrapolate the data at high temperatures with a linear fit,  as shown in Fig.~S2.
We find that the $f$ spectral weight reaches zero at  270$\pm$30~K (Fig.~S2), which is the estimated onset temperature of the localized-to-itinerant transition.

\newpage
\subsection*{Linear fits of the band dispersions near \boldmath{$E_F$}}

%%%%%%%%%%%%%%%%%%%%%%%%%%%% Figure S4 %%%%%%%%%%%%%%%%%%%%%%%%%%%%
\begin{figure}[htb]
\centering
\includegraphics[width=12cm]{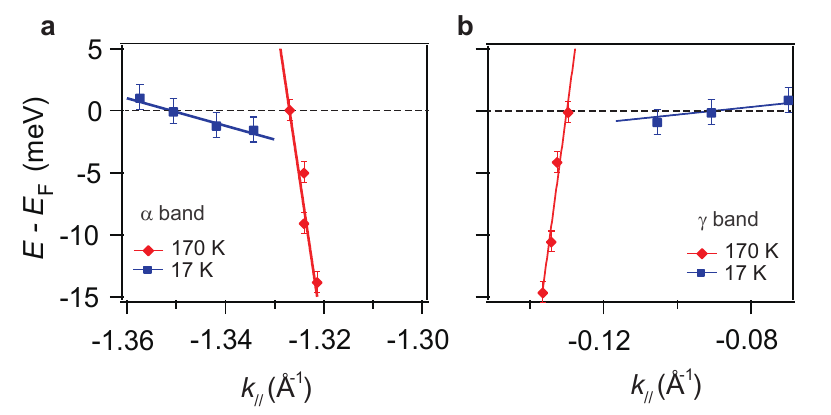}

\caption{
\textbf{a}, Linear fit of the electron-like $\alpha$ band centered at $M$ near $E_F$ for two  temperatures, which gives the dispersion relations:
$E(k)$[eV]=-2.677[eV$\cdot${\AA}]$(k-k_F)$[{\AA}$^{-1}$] at 170 K, and $E(k)$[eV]=-0.111[eV$\cdot${\AA}]$(k-k_F)$[{\AA}$^{-1}$] at 17 K.
 \textbf{b}, Linear fit of the electron-like $\gamma$ band centered at $\Gamma$ near $E_F$, which gives the dispersion relations:  $E(k)$[eV]=3.14[eV$\cdot${\AA}]($k-k_F$)[{\AA}$^{-1}$] at 170 K, and $E(k)$[eV]=0.031[eV$\cdot${\AA}]($k-k_F$)[{\AA}$^{-1}$] at 17 K.
}
\end{figure}
%%%%%%%%%%%%%%%%%%%%%%%%%%%%%%%%%%%%%%%%%%%%%%%%%%%%%%%%%%%%%%%%%%

Linear fits are conducted for the dispersions of the conduction bands ($\alpha$ and $\gamma$) near $E_F$ at 17 and 170~K. The high temperature fast dispersion was obtained  by tracking the MDCs. The dispersion data at 17 K are derived from the EDCs after dividing by the corresponding resolution-convoluted Fermi Dirac distribution (RC-FDD) (examples can be found in Fig.~S3 below). Due to the flatness of the band, weak spectral weight,  thermal broadening effects, and particularly the influence of the first excited CEF state, the dispersion data  in the intermediate temperature range cannot be accurately determined near $E_F$.
The Fermi velocity for the $\alpha$ band is 2.67\,eV$\cdot$\AA\ at 170 K, and it is reduced by $24 \pm 4$ times at 17 K (Fig. S3a), while for the $\gamma$ band, the Fermi velocity is 3.14\,eV$\cdot${\AA}\ at 170 K, and it is reduced by $102 \pm 14$ times at 17 K after the weakly dispersive hybridized band is formed (Fig. S3b).

%The conduction-band dispersion at high temperatures can be approximated as $\varepsilon(k)$[eV]=-2.677[eV$\cdot$\AA]($k-k_F)$[\AA$^{-1}$] for the $\alpha$
%band (Fig. S1A in SOM) and $\varepsilon(k)$[eV]=3.14[eV$\cdot$\AA]$(k-k_F)$[\AA$^{-1}$] for the $\gamma$ band (Fig. S2B in SOM).

\newpage
\subsection*{Parabolic fit of the large scale dispersion at  170 K}

%%%%%%%%%%%%%%%%%%%%%%%%%%%%% Figure. S5 %%%%%%%%%%%%%%%%%%%%%%%%%%%%
\begin{figure}[htb]
\centering
\includegraphics[width=8.7cm]{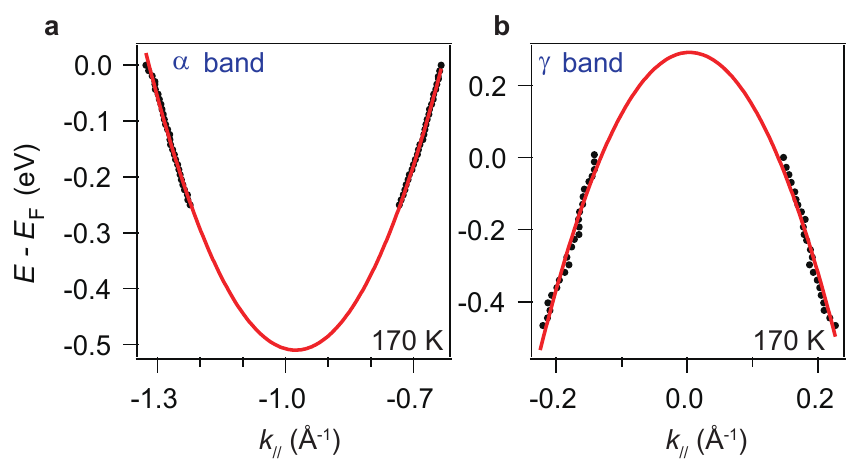}
\caption{
\textbf{Parabolic fit of the $\alpha$ and $\gamma$ band at 170 K. }
\textbf{a},\textbf{b} Fitting of the electron-like $\alpha$ band  centered at $M$ (\textbf{a}) and the hole-like $\gamma$ band centered at $\Gamma$ (\textbf{b}) by parabolic curves. The data were taken at 170 K. The circles represent the dispersions of the two bands obtained by fitting MDCs. The fitting curves are E$_k$=4.34$k^2$+8.47$k$+3.62 for $\alpha$, and E$_k$=-15.83$k^2$+0.12$k$+0.29 for $\gamma$, which gives m$^*$=0.88~m$_e$ for $\alpha$, and 0.24~m$_e$ for $\gamma$.}
\end{figure}
%%%%%%%%%%%%%%%%%%%%%%%%%%%%%%%%%%%%%%%%%%%%%%%%%%%%%%%%%%%%%%%%%%

In order to estimate the effective masses of the $\alpha$ and $\gamma$ bands at high temperature, we use parabolic curves to fit their high temperature dispersion (Fig.~S4). The relationship between E$_k$ and $k$ is E$_k$=4.34$k^2$+8.47$k$+3.62 for $\alpha$, and E$_k$=-15.83$k^2$+0.12$k$+0.29 for $\gamma$, which gives m$^*$=0.88~m$_e$ for $\alpha$ and 0.24~m$_e$ for $\gamma$.
Since the fit involves the dispersion over a large window of several hundreds of meV, where the $f$ electron contribution is negligible, the fitted value represents the effective mass of the corresponding band when $f$ electrons are completely localized at high temperatures. Moreover, we note that the temperature dependence of the experimental dispersions is very weak at high temperatures, when the $f$ electron contribution is negligible.

Based on this and the Fermi velocity renormalization obtained above, one can estimate the effective mass at 17 K in the coherent state to be about 21.1~$m_e$ for the $\alpha$ band, and 24.5~$m_e$ for the $\gamma$ band.

\newpage
\subsection*{Quantitative analysis of the enlargement of the \boldmath{$\alpha$} and \boldmath{$\gamma$}  bands}

%%%%%%%%%%%%%%%%%%%%%%%%%%%%% Figure. S6 %%%%%%%%%%%%%%%%%%%%%%%%%%%%
\begin{figure}
\centering
\includegraphics[width=13.5cm]{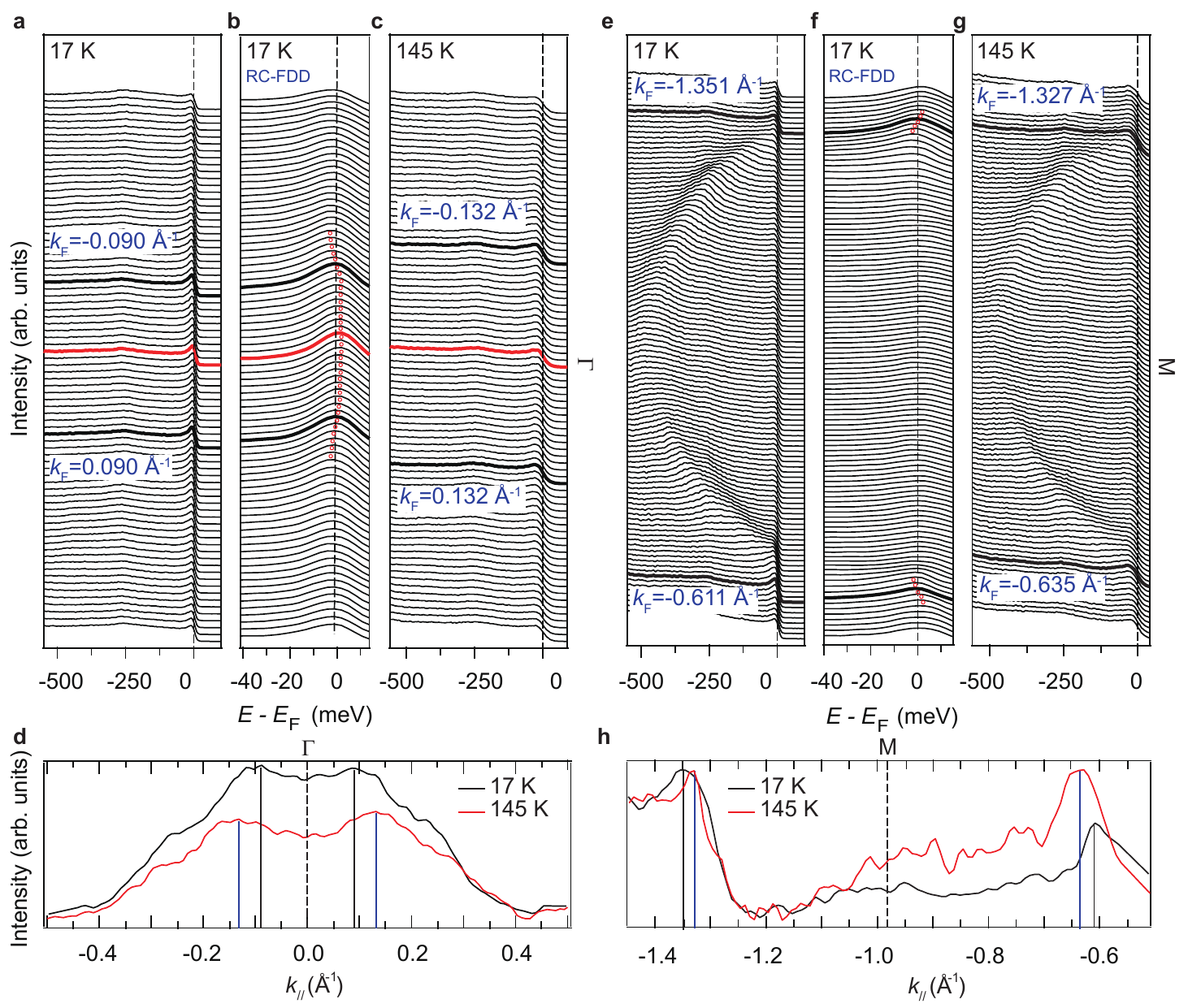}
\caption{
\textbf{Quantitative analysis of the enlargement of the $\alpha$ and $\gamma$ band. }
\textbf{a}, EDCs of the hole-like $\gamma$ band around $\Gamma$ taken at 17 K. \textbf{b}, EDCs of (\textbf{a}) divided by the RC-FDD in a smaller energy window to highlight the fine structure near $E_F$. The red circles mark the dispersion of the weakly dispersive hybridized band. A Fermi momentum ($k_F$) is determined based on the EDC whose peak is centered at $E_F$. The thick black curves denote the EDCs at $k_F$.  The thick red curves denote the EDCs at the $\Gamma$ point. \textbf{c}, EDCs of the $\gamma$ band in the same momentum region as (\textbf{a}), taken at 145 K. \textbf{d}, The MDCs at $E_F$ around $\Gamma$, taken at 17 and 145\,K. \textbf{e}, EDCs of the electron-like $\alpha$ band around $M$ taken at 17 K. \textbf{f}, EDCs of (\textbf{e}) divided by the RC-FDD in a smaller energy window to highlight the fine structures near $E_F$. \textbf{g}, EDCs of the $\alpha$ band in the same momentum region as (\textbf{e}), taken at 145 K. \textbf{h}, The  MDCs at $E_F$ around the $M$ point, taken at 17 and 145 K.
}\label{fig:FSexpansion}
\end{figure}
%%%%%%%%%%%%%%%%%%%%%%%%%%%%%%%%%%%%%%%%%%%%%%%%%%%%%%%%%%%%%%%%%%

The Fermi surface expands in the heavy electron state, due to the inclusion of additional $f$ electrons into the Fermi sea. Fig.~S5 presents a detailed quantitative analysis of the electron pocket around $M$ and the hole pocket around $\Gamma$.
Figs.~S5a and S5b plot the EDCs of the $\gamma$ band taken at 17 K along $\Gamma$-$M$ and the corresponding EDCs after dividing by the RC-FDD, respectively.
The dispersion of the hybridized band is clearly observed in
Fig. S5b, and the Fermi momentum at low temperature can be determined when the peak passes $E_F$, which is labeled by the thick black lines. Alternatively, the same Fermi momenta could be obtained through the MDC maxima (Fig. S5d). At high temperature, the $k_F$s could be determined from the MDC maxima. Consequently, the radius of the $\gamma$ hole pocket is estimated to be $0.042\pm0.008$\AA\ smaller at 17 K for the $\gamma$ band, while for the $\alpha$ band, the radius of this electron pocket is estimated to be $0.024\pm0.008$\AA\ larger at 17 K.  Considering the rather two-dimensional character of the $\alpha$ band and assuming an isotropic expansion, the enlargement indicates that about 0.058 electrons have been transferred from the local $f$ moment to $\alpha$ pocket at low temperature.

\newpage
\subsection*{CEF excitations and the temperature dependence of the CEF level occupation}

The Kondo energy scale $T_K$ increases exponentially with the degeneracy of the 4$f$ ground state.
Such a case would correspond to a vanishing CEF excitation gap $\Delta$. On the other hand, for sufficiently large splitting $\Delta$ the effect of the doublet on the Kondo energy scale is negligible.
It  has been well established that the presence of CEF excitations with the gaps of the order of $T_K$ can also result in a considerable enhancement of the low-energy scale~\cite{Cornut.72, Kroha.03}.
Our results demonstrate that even in CeCoIn$_5$, where the common belief has been that CEF excitations only have a weak influence on the low-energy state, the CEF excitation gaps lead to a fine structure in the Kondo resonance. The onset of heavy band formation is reflected in the dispersion  at temperatures well above the coherence temperature, {\it i.e.} the maximum in the resistivity.
Our results suggest that the role played by CEF levels goes beyond the simple enhancement of the Kondo temperature.

Our observation of the Kondo satellite peaks at energies $E_{2}=9$ meV and $E_{3}=30$ meV is in good agreement with the findings of Bauer {\it et al}.~\cite{Bauer.04} and confirm their analysis.
Following Bauer {\it et al}., the CEF doublets in the $|J_z\rangle$ basis of the $J=5/2$ state are
\begin{eqnarray}
\label{eq:S1}
\Gamma_7^{(1)} &=& a |\pm 5/2\rangle -b |\mp 3/2 \rangle \nonumber\\
\Gamma_7^{(2)} &=& b |\pm 5/2\rangle +a |\mp 3/2 \rangle\\
\Gamma_{6} &=& |\pm 1/2 \nonumber \rangle
\end{eqnarray}

The widths and heights of the additional Kondo satellite peaks strongly depend on the elements of the hybridization matrix, {\it i.e.} on the various overlaps between the crystal field states and the local conduction electron states.
%The comparatively weak feature of the first excited CEF doublet as compared to the second excited CEF doublet suggests that the corresponding hybridization matrix elements are larger. (DLFeng comment, we cannot be sure if this is not artifact of the division)
From the characteristic shapes of the orbitals in Eq.~(\ref{eq:S1}) it seems natural to expect that the two-dimensional $\alpha$ band will have larger overlap with $\Gamma_7^{(1)}$ and $\Gamma_7^{(2)}$ than $\Gamma_6$.
This expectation would be in line with the observed smaller mass enhancement of the $\alpha$ band as compared to the $\gamma$ band. DMFT+LDA might be able to confirm this expectation.
Moreover, estimating the  enhancement factor of the Kondo energy scale by CEF effects requires a knowledge of the CEF splittings and the full hybridization matrix that describes the coupling between the 4$f$ and conduction electrons~\cite{Kroha.03}. This also requires further DMFT+LDA calculations.

Based on the energies obtained for the two CEF excitation gaps, we can estimate the average occupation of each of the doublets under the assumption that the valence of Ce in CeCoIn$_5$ remains 3+, independent of temperature for the temperature range considered here.

Although the $f$-level occupation in the presence of a non-vanishing hybridization is not a good quantum number in the Anderson model, the average occupation in the single-level Anderson impurity model is only a weak function of temperature. This allows us to estimate the average occupation of the CEF split levels  in the local moment regime, where $n_0+n_1+n_2\rightarrow 1$.
Similarly, the Ce valence in CeCoIn$_5$ is only weakly temperature-dependent and we can estimate the average occupation of the CEF split levels in the $4f_{5/2}$ multiplet using
\begin{equation}
\label{eq:n}
\langle n_i \rangle=\frac{e^{-\Delta_i/T}}{1+e^{-\Delta_1/T}+e^{-\Delta_2/T}}~(i=1,2).
\end{equation}
Here, $\Delta_1=7$ meV and $\Delta_2=28$ meV are the CEF splittings inferred from Fig.~3c of the main text. $n_1$ ($n_2$) represents the average occupation of the first (second) excited CEF level. As shown in Fig.\ 5b, $n_1$ changes considerably in the temperature window of our experiment while the occupation of the highest CEF level (dashed line in Fig.\ 5b) can be neglected.  Interestingly, the slope of Eq.\ (\ref{eq:n}) for $i=1$ changes at $T=20$ K and more smoothly at around $T=90$ K  which corresponds to the temperatures where the spectrum near the Fermi energy changes, see Fig.\ 5a. It would be interesting to analyze optical conductivity measurements in terms of $E/T$ scaling and its implications for vertex corrections.

\newpage
\subsection*{Implications for YbRh$_2$Si$_2$}

Our results may also shed new light on the apparent temperature independence of the Fermi surface in YbRh$_2$Si$_2$ reported in Ref.~\cite{Kummer.15}.
The CEF excitations of YbRh$_2$Si$_2$ have been discussed in Ref.~\cite{Stockert.06}. The concomitant Kondo satellite peaks in YbRh$_2$Si$_2$ have been seen in STM measurements \cite{Ernst.11}, in good agreement with \cite{Stockert.06}: $\Delta_1^{\mbox{\tiny YRS}}=17$ meV, $\Delta_2^{\mbox{\tiny YRS}}=27$ meV and $\Delta_3^{\mbox{\tiny YRS}}=43$ meV.
%The higher number of CEF splittings compared to CeCoIn$_5$ is a consequence of the $4f_{7/2}$ state.
The analysis of Yb-based compounds is complicated by the hole-character of the 4$f$-state {\it i.e.} the large number of electrons in the 4$f$-shell. Nonetheless, it seems reasonable to expect that the apparent temperature independence reported in \cite{Kummer.15} is related to the larger CEF gaps $\Delta_1^{\mbox{\tiny YRS}}=17$~meV~$\approx 200$\,K of YbRh$_2$Si$_2$ as compared to CeCoIn$_5$. It thus seems imperative to extend the temperature range studied in \cite{Kummer.15} to higher temperatures.


\begin{thebibliography}{10}
%\begin{references}


\bibitem{Varma.76} Varma, C. M. et al. Mixed-valence compounds. ${Rev.~Mod.~Phys.}$ \textbf{48}, 219 (1976).
%\bibitem{Tahvildar-Zadeh.98} Tahvildar-Zadeh, A. N. et al. Low-temperature coherence in the periodic Anderson model: predictions for photoemission of heavy fermions. ${Phys.~Rev.~Lett.}$ \textbf{80}, 5168 (1998).
\bibitem{Stewart.84} Stewart, G. R. et al. Heavy-fermion systems. ${Rev.~Mod.~Phys.}$ \textbf{56}, 755 (1984).
%\bibitem{Allen.05} Allen, J. W. et al. The Kondo resonance in electron spectroscopy. ${J.~Phys.~Soc.~Jpn.}$ \textbf{74}, 34 (2005).
\bibitem{Shim.07} Shim, J. H. et al. Modeling the localized-to-itinerant electronic transition in heavy fermion system CeIrIn$_5$. ${Science}$ \textbf{7318}, 1615 (2007).
\bibitem{Choi.12} Choi, H. C. et al. Temperature-dependent Fermi surface evolution in heavy fermion CeIrIn$_5$. ${Phys.~Rev.~Lett.}$ \textbf{108}, 016402 (2012).
\bibitem{Mo.12} Mo, S. K. et al. Emerging coherence with unified energy, temperature, and lifetime scale in heavy fermion YbRh$_2$Si$_2$. ${Phys.~Rev.~B}$ \textbf{85}, 241103(R) (2012).
\bibitem{Kummer.15} Kummer, K. et al. Temperature-independent Fermi surface in the Kondo lattice YbRh$_2$Si$_2$. ${Phys.~Rev.~X}$ \textbf{5}, 011028 (2015).
\bibitem{Fujimori.07} Fujimori, S. et al. Itinerant to localized transition of $f$ electrons in the antiferromagnetic superconductior UPd$_2$Al$_3$. ${Nature.~Phys.}$ \textbf{3}, 618 (2007).

\bibitem{Shishido.02} Shishido, H. et al. Fermi Surface, Magnetic and Superconducting Properties of LaRhIn$_5$ and CeTIn$_5$ (T: Co, Rh and Ir). ${J.~Phys.~Soc.~Jpn.}$ \textbf{71}, 162-173 (2002).
\bibitem{Settai.01} Settai. R, et al. Quasi-two-dimensional Fermi surfaces and the de Haas-van Alphen oscillation in both the normal and superconducting mixed states of CeCoIn$_5$. ${J.~Phys.:Condens.~Matter}$ \textbf{13}, L627 (2001).
\bibitem{Singley.02} Singley, E. J. et al. Optical conductivity of the heavy fermion superconductor CeCoIn$_5$. ${Phys.~Rev.~B}$ \textbf{65}, 161101(R) (2002).
%\bibitem{Mena.05} Mena, F. P. et al. Optical conductivity of CeMIn$_5$ (M= Co, Rh, Ir). ${Phys.~Rev.~B}$ \textbf{72}, 045119 (2005).
\bibitem{Burch.07} Burch, K. S. et al. Optical signatures of momentum-dependent hybridization of the local moments and conduction electrons in Kondo lattices. ${Phys.~Rev.~B}$ \textbf{75}, 054523 (2007).
\bibitem{Aynajian.12} Aynajian, P. et al. Visualizing heavy fermions emerging in a quantum critical Kondo lattice. ${Nature}$ \textbf{486}, 201 (2012).
\bibitem{Allan.13} Allan, M. P. et al. Imaging Cooper pairing of heavy fermions in CeCoIn$_5$. ${Nature.~Phys.}$ \textbf{9}, 468 (2013).
\bibitem{Zhou.13} Zhou, B. B. et al. Visualizing nodal heavy fermion superconductivity in CeCoIn$_5$. ${Nature.~Phys.}$ \textbf{9}, 474 (2013).
\bibitem{Haule.10} Haule, K. et al. Dynamical mean-field theory within the full-potential methods: Electronic structure of CeIrIn$_5$, CeCoIn$_5$, and CeRhIn$_5$. ${Phys.~Rev.~B}$ \textbf{81}, 195107 (2010).
\bibitem{Koitzsch.13} Koitzsch, A. et al. Band-dependent emergence of heavy quasiparticles in CeCoIn$_5$. ${Phys.~Rev.~B}$ \textbf{88}, 035124 (2013).
\bibitem{Koitzsch.08} Koitzsch, A. et al. Hybridization effects in CeCoIn$_5$ observed by angle-resolved photoemission. ${Phys.~Rev.~B}$ \textbf{77}, 155128 (2008).
\bibitem{Koitzsch.09} Koitzsch, A. et al. Electronic structure of CeCoIn$_5$ from angle-resolved photoemission spectroscopy. ${Phys.~Rev.~B}$ \textbf{79}, 075104 (2009).
\bibitem{Paschen.16} Paschen, S. et al. Kondo destruction in heavy fermion quantum criticality and the photoemission spectrum of YbRh$_2$Si$_2$. ${J.~ Magn.~Magn.~Mater.}$ \textbf{400}, 17-22 (2016).
\bibitem{Petrovic.01} Petrovic, C. et al. Heavy-fermion superconductivity in CeCoIn$_5$. ${J.~Phys.:Condens.~Matter}$ \textbf{13}, L337 (2001).
\bibitem{Bel.04} Bel, R. et al. Giant Nernst in CeCoIn$_5$. ${Phys.~Rev.~Lett.}$ \textbf{92}, 217002 (2004).
\bibitem{Strocov.12} Strocov, V. N. et al. Three-Dimensional Electron Realm in VSe$_2$ by Soft-X-Ray Photoemission Spectroscopy: Origin of Charge-Density Waves. ${Phys.~Rev.~Lett.}$ \textbf{109}, 086401 (2012).
\bibitem{Pfleiderer.09} Pfleiderer, C. et al. Superconducting phases of $f$-electron compounds. ${Rev.~Mod.~Phys.}$ \textbf{81}, 1551 (2009).
\bibitem{Fujimori.03} Fujimori, S. -i. et al. Nearly localized nature of $f$ electrons in CeTIn$_5$ (T= Rh, Ir). ${Phys.~Rev.~B}$ \textbf{67}, 144507 (2003).
\bibitem{Fujimori.06} Fujimori, S. -i. et al. Direct observation of a quasiparticle band in CeIrIn$_5$: An angle-resolved photoemission spectroscopy study. ${Phys.~Rev.~B}$ \textbf{73}, 224517 (2006).
\bibitem{Kroha.03} Kroha, J. et al. Structure and transport in multi-orbital Kondo systems. ${Physica~E}$ \textbf{18}, 69-72 (2003).
\bibitem{Bauer.04} Bauer, E. D. et al. Crystalline electric field excitations in the heavy fermion superconductor CeCoIn$_5$. ${J.~Appl.~Phys.}$ \textbf{95}, 7201 (2004).
\bibitem{footnote1} We note that the heights of the satellite CEF peaks appear stronger than that of the main one at high temperatures, which may be due to matrix element effects or a divergence artifact from the division by RC-FDD high above $E_F$.
\bibitem{Cornut.72} Cornut, B. et al. Influence of the crystalline field on the Kondo effect of alloys and compounds with cerium impurities. ${Phys.~Rev.~B}$ \textbf{5}, 4541 (1972).
\bibitem{Daou.09} Daou, R. et al. Linear temperature dependence of resistivity and charge in the Fermi surface at the pseudogap critical point of a high-T$_c$ superconductor. ${Nature.~Phys.}$ \textbf{5}, 31-34 (2009).
\bibitem{Dai.13} Dai, Y. M. et al. Hidden T-linear scattering rate in Ba$_{0.6}$K$_{0.4}$Fe$_2$As$_2$ revealed by optical spectroscopy. ${Phys.~Rev.~Lett.}$ \textbf{111}, 117001 (2013).
\bibitem{Bruin.13} Bruin, J. A. N. et al. Similarity of scattering rates in metals showing T-linear resistivity. ${Science}$ \textbf{339}, 804 (2013).
\bibitem{Valla.99} Valla, T. et al. Evidence of quantum critical behavior in the optimally doped cuprate Bi$_2$Sr$_2$CaCu$_2$O. ${Science}$ \textbf{285}, 2110 (1999).
\bibitem{Hewson.93} A. C. Hewson, The Kondo Problem to Heavy Fermions (Cambridge University Press, Cambridge, 1993).
\bibitem{Im.08} Im, H. J. et al. Direct observation of dispersive Kondo resonance peaks in a heavy-fermion system. ${Phys.~Rev.~Lett.}$ \textbf{100}, 176402 (2008).
\bibitem{Hall.01} Hall, D. et al. Fermi surface of the heavy-fermion superconductor CeCoIn$_5$: The de Haas-van Alphen effect in the normal state. ${Phys.~Rev.~B}$ \textbf{64}, 212508 (2001).
\bibitem{Booth.11} Booth, C. H. et al. Electronic structure and $f$-orbital occupancy in Yb-substituted CeCoIn$_5$. ${Phys.~Rev.~B}$ \textbf{83}, 235117 (2011).
\bibitem{Treske.14} Treske, U. et al. X-ray photoemission study of CeTIn$_5$ ${J.~Phys.~Soc.~Jpn.}$ \textbf{26}, 205601 (2014).
\bibitem{Varma.89} Varma, C. M. et al. Phenomenology of the normal state of Cu-O high-temperature superconductors. ${Phys.~Rev.~Lett.}$ \textbf{63}, 1996 (1989).
\bibitem{Varma.16} Varma, C. M. et al. Quantum-critical flucuations in 2D meatals: strange metals and superconductivity in antiferromagnets and in cuprates. ${Rep.~Prog.~Phys.}$ \textbf{48}, 219 (2016).
\bibitem{Aronson.95} Aronson, M. C. et al. Non-Fermi-liquid scaling of the magnetic response in UCu$_{5-x}$Pd$_x$ (x=1, 1.5). ${Phys.~Rev.~Lett.}$ \textbf{75}, 725 (1995).
\bibitem{Schroder1} Schr\"oder, A. et al. Scaling of magnetic fluctuations near a quantum phase transition. ${Phys.~Rev.~Lett.}$ \textbf{80}, 5623 (1998).
\bibitem{Friedemann.10} Friedemann, S. et al. Fermi-surface and dynamical scaling near a quantum-critical point. ${Proc.~Natl.~Acad.~Sci.~USA}$ \textbf{107}, 14547 (2010).
\bibitem{footnote2} Such an $E/T$ scaling could be observed at momenta where the $f$ states dominate the spectra
near $E_F$ . To obtain the single particle spectral function from the data shown in Fig. 3b,
the data need to be divided by the Fermi Dirac distribution function. This, however, is a
function of $E/T$ only and cannot spoil the scaling properties shown in Fig. 5b.
\bibitem{footnote3} It may be worthwhile to investigate if and how the $E/T$ scaling observed here relates to
the quantum critical fluctuation spectrum observed in CeCoIn$_5$ at much lower temperatures
and to that of related systems.
\bibitem{Stockert.12} Stockert, O. et al. Superconductivity in Ce- and U-based "122" heavy fermion compounds. ${J.~Phys.~Soc.~Jpn.}$ \textbf{81}, 011001 (2012).
\bibitem{Ernst.11} Ernst, S. et al. Emerging local Kondo screening and spatial coherence in the heavy-fermion metal YbRh$_2$Si$_2$. ${Nature}$ \textbf{474}, 362 (2011).
\bibitem{Stockert.06} Ernst, S. et al. Crystalline electric field excitations of the non-Fermi-liquid YbRh$_2$Si$_2$. ${Physica B}$ \textbf{378-380}, 157 (2006).

%\end{references}
\end{thebibliography}
\end{document}